\documentclass{aastex63}

\usepackage{color,amsmath}
\newcommand\Msun{\mbox{$M_\sun$}}
\newcommand\Mspy{\mbox{\Msun\,yr$^{-1}$}}
\newcommand\kmps{\mbox{km\,s$^{-1}$}}

\shorttitle{CW Leo Wind Forming Region}
\shortauthors{Kim et al.}

\begin{document}
\title{Near-infrared Integral-field Spectroscopy of the Wind Forming Region of CW Leo}

\author[0000-0001-9639-0354]{Hyosun Kim}
\affiliation{Korea Astronomy and Space Science Institute, 776 Daedeok-daero, Yuseong-gu, Daejeon 34055, Republic of Korea}

\author[0000-0001-9490-3582]{Youichi Ohyama}
\affiliation{Institute of Astronomy and Astrophysics, Academia Sinica, 11F of Astronomy-Mathematics Building, AS/NTU, No.1, Sec. 4, Roosevelt Rd, Taipei 106319, Taiwan, R.O.C.}

\author[0000-0002-3808-7143]{Ho-Gyu Lee}
\affiliation{Korea Astronomy and Space Science Institute, 776 Daedeok-daero, Yuseong-gu, Daejeon 34055, Republic of Korea}

\author[0000-0002-1418-3309]{Ji Hoon Kim}
\affiliation{SNU Astronomy Research Center, Astronomy Program, Department of Physics and Astronomy, Seoul National University, 1 Gwanak-Ro, Gwanak-Gu, Seoul 08826, Republic of Korea}

\correspondingauthor{Hyosun Kim}
\email{hkim@kasi.re.kr}

\begin{abstract}
The circumstellar envelope of the carbon star CW Leo exhibited various unexpected changes in recent optical imaging observations. We have performed a follow-up observation using the Near-infrared Integral-Field Spectrograph (NIFS) equipped on the Gemini-North telescope.
We report the near-infrared counterparts of a local brightness peak in the optical at the stellar position of CW Leo. On the other hand, a second peak detected at short wavelengths in the $J$ band coincides with the brightest, bluest position in the optical images.
The absorption features in the $K$ band are minimized at a radius of 0\farcs2 from the predicted stellar position. The reduction of the absorption depths likely indicates dilution of the absorption features by thermal emission of dust grains newly formed at such a radius and heated by radiation from the central star.
The broad absorption feature at 1.53\,\micron\ is significantly deeper than in template carbon stars, consistent with the presence of a substantial amount of circumstellar material around CW Leo.
Its northeastern quadrant lacks circumstellar absorption features and scattered light in the near-infrared regime, which are possibly manifestations of its conical cavity in both gas and dust.
In addition, a cross correlation of CO overtone bands indicates that the average expansion velocity of dust grains is smaller to the northern direction, likewise the velocity of transverse wind components derived using the differential proper motion of a circumstellar whirled pattern.
The gradual brightening of CW Leo and the changes in its innermost circumstellar envelope need further continuous monitoring observations to properly understand its transitional phase toward the post-asymptotic-giant-branch stage.
\end{abstract}


\section{Introduction}\label{sec:int}

CW Leo (aka IRC+10216) is one of the most well-known carbon stars (aka carbon-rich asymptotic giant branch (AGB) stars). Thanks to its proximity locating at 123\,$\pm$\,14\,pc \citep{gro12} and high mass-loss rate of $\ga\,2\times10^{-5}\,\Mspy$ \citep[e.g.,][]{gue18}, it has been favored for various observations in all wavelengths as a laboratory for studying circumstellar chemical elements, dust, and geometry of the wind \citep[e.g.,][]{gue77,jur83,cra87,mau99,fon03,dec11,cer15,gue18,and24,tuo24}.

One of the most important questions in the studies of late stellar evolution is related to the origin of shape transition from nearly spherical AGB winds to highly aspherical (polar) outflows found in planetary nebulae (PNe), and the only consensus in the community is that the companion(s) must be involved in the shaping process \citep[e.g.,][]{dem09}. CW Leo is a good test bed for understanding the shaping of the wind, albeit mild as it is still in a progenitor stage of a PN, by tracing the circumstellar material and the characteristic pattern as the footprints of unseen companion(s), again thanks to its proximity, which enables us to achieve observations at a very high spatial resolution. Since the discovery of its multiple shells enshrouding the carbon star \citep{mau99}, the presence of a companion star has been continuously predicted from a theoretical basis due to a spiral-shell pattern it creates \citep{mas99,kim12}, the whorled pattern's molecular line emissions in S-shaped position-velocity diagrams \citep{dec15}, and the finding of an equatorial dust lane \citep{jef14}. Furthermore, even a second companion is cautiously suggested in order to account for the complexity of the circumstellar pattern including the off-centering of the approximated rings and the unforeseen abrupt radial change in the interval of the pattern \citep{kim24}.

CW Leo is getting brighter and this trend has continued for two decades \citep{kim15,kim21,kim24iau}. A clear and linear brightening trend of CW Leo was first reported by \citet{kim15} using photometric data of the Catalina Sky Survey performed since 2005 with unfiltered images \citep{dra14}; the trend lasts at least a decade, resulting in a total magnitude increase of $\sim1.6$\,mag, which is similar to an amplitude of $\sim1.8$\,mag of a sinusoidal variation due to a 640 day stellar pulsation. Similar brightening trends are manifested in the near- to mid-infrared $JHKLM$ bands in the light curves taken by Shenavrin (private communication) from 2017 up to date \citep[Figure\,6 of][]{kim24iau}, while those light curves taken in 1999--2010 by the same researchers using the same instruments did not display any brightening trends \citep{she11}.

The appearance of the recent brightening tendency of CW Leo may not be the first. As being reanalyzed by \citet[see their Figure\,2]{kim15}, the previous light curves reported by \citet{leb92} presented some brightening in the near-infrared $JHK$ bands for 4 yr (1985--1989), while no brightening in the mid-infrared $LM$ bands were reconfirmed as in the original literature. But because of the paucity of data points and the short duration of the photometric monitoring (only two pulsation periods), the reliability for the linear brightening rates in this old epoch may not be firmly concluded.

We hypothesize that CW Leo may undergo precursory events that are getting stronger and longer, prior to the eventual ejection of nearly its entire stellar mass. It is unclear whether the current brightening is the very final, single event immediate before ending the AGB phase, or just one event of a series of repeated events. The indisputable points include that the current brightening lasts unhistorically long and is global in a wide range of wavelengths from optical to mid-infrared. The high brightening rate of CW Leo is beyond the extinction reduction model after the cessation of stellar mass loss \citep{kam20}. 

The central carbon star of CW Leo seems to have recently revealed itself in the very central region of an optical image taken with the Hubble Space Telescope (HST) in 2016 \citep{kim21}\footnote{\dataset[https://doi.org/10.17909/5zs0-k163]{https://doi.org/10.17909/5zs0-k163}}. The same-filtered HST images taken in 1998 and 2001 presented a bipolar-like structure with the predicted (proper-motion-corrected) stellar position included within the brighter blob of the two (anticipated) bipolar blobs, while the same-filtered HST image taken in 2011 showed several smaller clumps surrounding the unoccupied stellar position and, surprisingly, the disappearance of the long-standing bipolar-like nebula (\citealp{kim15}; see also Figure 1 of \citealp{kim23iau}). Whether the emergence of the very red, compact, optical peak exactly at the stellar position indicates an unveiling of the star between 2011 and 2016, needs to be confirmed with equivalently high spatial resolution images at longer wavelengths. In this paper, we report the spectro-imaging data in the near-infrared bands, using integral-field unit (IFU) spectroscopy, taken in 2016 with an epoch difference of less than a few months from the epoch of the HST image. In this paper, the near-infrared counterpart of the optical, point-like local brightness peak is firmly identified for the first time. The spatial distribution of the central circumstellar envelope is also examined with a special interest toward the direction of the previously known bipolar cone.

This paper is organized as follows. In Section\,\ref{sec:obs}, the observations and data reduction are described. The results are shown in Sections\,\ref{sec:spa} and \ref{sec:spe}, focusing on the brightness center (Section\,\ref{sec:ast}), the brightness distribution (Section\,\ref{sec:dst}), the color maps (Section\,\ref{sec:clr}), the spectroscopic indices of absorption features and their spatial distributions (Sections\,\ref{sec:ind} and \ref{sec:ssd}), and the net velocities of dust grains measured through cross correlation of CO overtone bands (Section\,\ref{sec:exp}). A summary and implications of our findings are presented in Section\,\ref{sec:dis}.

\section{Observation and Data Reduction}\label{sec:obs}
\subsection{Gemini Observations}\label{sec:gem}

We performed the near-infrared spectro-imaging observations of the closest carbon star CW Leo using the Near-infrared Integral-Field Spectrograph \citep[NIFS;][]{mcg03} with the ALTtitude conjugate Adaptive optics for the InfraRed (ALTAIR) for the laser guide star (LGS) adaptive optics correction with the peripheral wave front sensor (PWFS1 or P1) of the Gemini-North telescope (Program ID: GN-2016A-Q-18). The angular resolution of LGS\,$+$\,P1 observing mode that we used is not diffraction limited but is instead given as ``superseeing,'' which is better than the natural seeing by a factor of 2--3. This observing mode was the best option for CW Leo, because no available guiding star exists within 25\arcsec\ for ALTAIR natural guide star (NGS) or LGS mode and guiding on source (as it is sufficiently bright) may require very careful consideration of the extended distribution ($>1\arcsec$) of the circumstellar envelope of CW Leo.

We obtained $J$-, $H$-, and $K$-band data on 2016 February 24, 28, and April 18, respectively. The average air masses during the target observations were 1.01, 1.06, and 1.36 for $J$, $H$, and $K$ bands, respectively. NIFS provided a field of view of $3\arcsec\times3\arcsec$ and a pixel scale of $0\farcs103\times0\farcs04$ with the larger value across slices. We set the individual exposure times for the $J$, $H$, and $K$ bands to 100, 10, and 22\,s, respectively, avoiding saturation, with the addition of the KG3 neutral density (ND) filter with transmission of 5\% in $K$ band. With 2, 8, and 8 coadds and 6, 4, and 6 on-source frames between blank sky frames, the resulting total on-source times are 1200, 320, 1056\,s for $J$, $H$, and $K$ bands, respectively.

We observed the standard star HIP~56736 (A0\,V, $V=8.8$\,mag) for flux and telluric calibrations at a similar air mass immediate before, or after, the target observations. The average air masses during the standard star observations were 1.33, 1.08, and 1.00 for $J$, $H$, and $K$ bands, respectively.

\subsection{Data Reduction}\label{sec:red}
For each dataset of the science and standard star frames, we reduced and calibrated them following standard procedures discussed on the Gemini-NIFS Data Reduction web page\footnote{https://www.gemini.edu/instrumentation/nifs/data-reduction}. We used PyRAF scripts (\verb+NIFS_Basecalib.py+, \verb+NIFS_Telluric.py+, and \verb+NIFS_Science.py+) running on Gemini IRAF package\footnote{https://www.gemini.edu/observing/phase-iii/reducing-data/gemini-iraf-data-reduction-software} version 1.13 (implemented on IRAF\footnote{IRAF is distributed by the National Optical Astronomy Observatories, which is operated by the Association of Universities for Research in Astronomy, Inc. (AURA) under cooperative agreement with the National Science Foundation.} version 2.16). The standard procedures included flat-field correction for both the pixel-to-pixel sensitivity variation and slit functions over 29 slitlets of the IFU, subtraction of the sky spectrum using the dedicated sky exposures, wavelength calibration using the Ar and Xe arc lamp spectrum, spatial distortion correction using the Ronchi flat exposures, and reconstruction of spectral cubes. The final data cubes for individual exposures are resampled to an angular pixel size of $0\farcs05\times0\farcs05$.
For $K$ band, we additionally measured the ND filter response using the flat-field images taken with and without the ND filter, and the science frames were divided by this response to remove the attenuation by the ND filter. For each band, the individual frames were aligned by locating the brightness maxima at the origin of coordinates. Finally, these 6, 4, and 6 individual cubes for $J$, $H$, and $K$ bands, respectively, were median averaged to generate the final cubes.

We further elaborated the telluric correction, spectral response calibration, and flux calibration by modifying the procedures implemented in the above Gemini PyRAF scripts. Good flux calibration is required for good flux consistency among the three bands, and careful removal of intrinsic stellar absorption of the A-type star spectrum is required to reliably measure the absorption features of CW Leo. To build the intrinsic spectrum of the standard star, we assumed that the star is a blackbody with a temperature appropriate for an A0\,V-type star (9600\,K) and used Two Micron All-Sky Survey photometry of the star for the absolute-flux calibration. Hydrogen absorptions in both the Paschen and Brackett series were fit assuming a Voigt profile and removed from the observed spectra of the standard star. In particular, we paid special attention to the secure removal of even fainter absorptions in the $H$ band through a simultaneous fit for multiple higher-order Brackett lines.

The reconstructed point-spread-function sizes were measured as the full width at half maximum (FWHM) of the cubes for the standard star (observed in the NGS mode), resulting in 0\farcs1--0\farcs2, depending on the bands under different average air masses. We regard that the angular resolution of our target observations in the LGS\,$+$\,P1 (superseeing) mode is slightly worse than these values. Notice that the stellar radius of CW Leo ($R_* \sim 0\farcs015$, according to \citealp{men12}; see also \citealp{deb12} and references therein, suggesting $\sim0\farcs02$) is unresolved in the NIFS observations, while the wind acceleration zone \citep[$\la0\farcs2$ in radius, based on various molecular line observations of][]{dec15} is resolvable.

The resolving powers ($R=\lambda/d\lambda$) measured using the lines of the Ar and Xe arc lamps were 7460, 5620, and 5370 for the $J$, $H$, and $K$ bands, respectively, corresponding to a velocity resolution of $\ga40\,\kmps$. The relative wavelength (or velocity) calibration accuracy across the entire field of view of NIFS, on the other hand, was estimated to be 1.4\,\kmps\ ($1\sigma$) in $K$ band, by analyzing the cubes of the arc lamps processed exactly in the same way for both the wavelength calibration and the cube generation, as for the corresponding cube for the target.

We note that each cube for each band was observed on different nights and was centered independently during target acquisition. Therefore, spatial registration of the cubes relies on some assumptions of the color-dependent morphology of CW Leo (Section\,\ref{sec:ast}). We also note that the spectral absolute-flux calibration was independently made on each cube. This, together with the spatial registration error, may cause small jumps in flux between the bands.

\section{SPATIAL DISTRIBUTION}\label{sec:spa}
\subsection{Identification of the Brightest Peak and Image Alignments}\label{sec:ast}
\begin{figure*} 
  \epsscale{1.15}
  \plotone{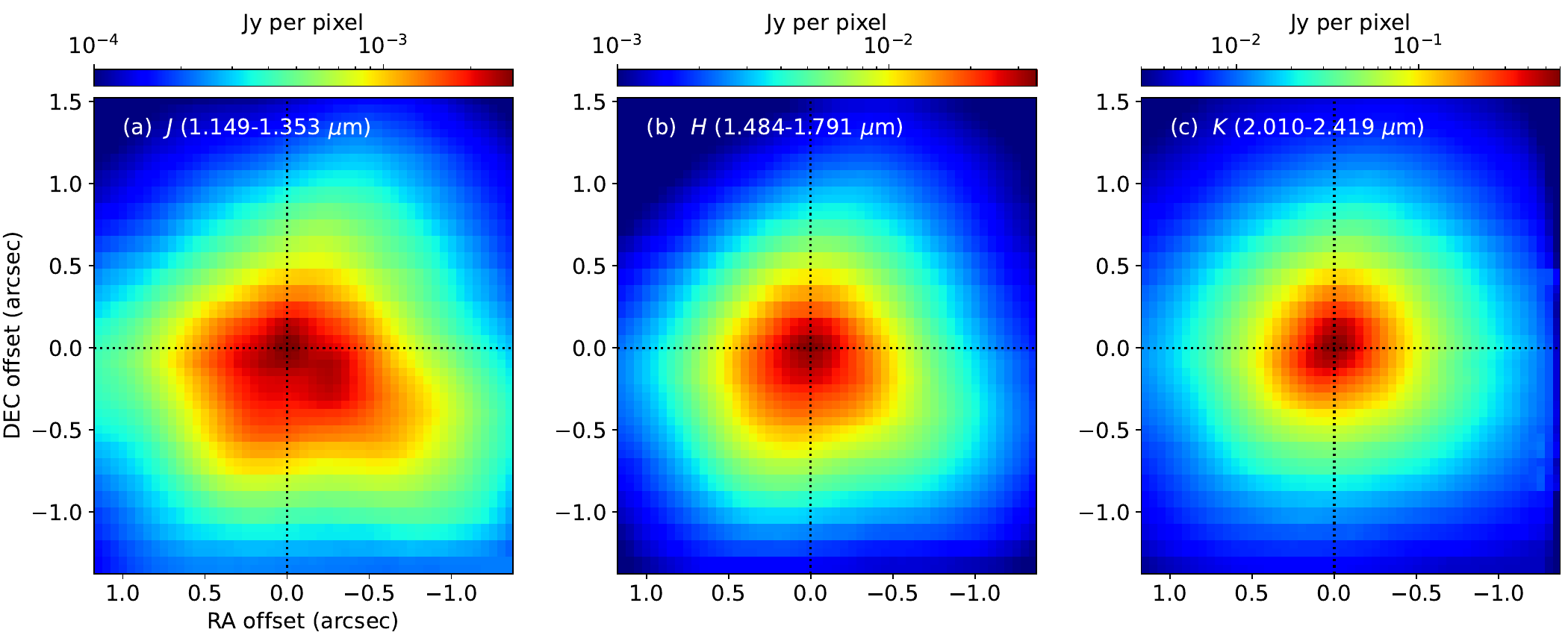}
  \caption{\label{fig:imgs}
    Near-infrared brightness distributions obtained from median-based
    averages of the Gemini-NIFS spectra over the wavelengths within the
    $J$, $H$, and $K$ bands. The color bars are shown in logarithmic
    scale in 0.0001--0.0028\,Jy\ pixel$^{-1}$, 0.001--0.035\,Jy\
    pixel$^{-1}$, and 0.003--0.6\,Jy\ pixel$^{-1}$, respectively,
    with a regridded pixel size of $0\farcs05\times0\farcs05$. Dotted
    lines indicate at their cross the brightest position in each band.
  }
\end{figure*}

Figure\,\ref{fig:imgs} exhibits the integrated brightnesses as being median averaged over wavelengths within the individual $J$, $H$, and $K$ bands with the image centers at their brightness maxima. Half of the image scale, $\sim 1\farcs5$, corresponds to $\sim 100 R_*$.
The brightness distributions in short wavelengths in the $J$-band data cube are double peaked with similar brightnesses between the two peaks. The secondary peak in the integrated $J$ brightness image, located at $\sim\,0\farcs4$ to the southwest, is progressively weakened along wavelength with respect to the primary peak. As a continuous trend, single-peaked brightness distributions are observed throughout $H$ and $K$ wavelengths.

\begin{figure*} 
  \epsscale{1.15}
  \plotone{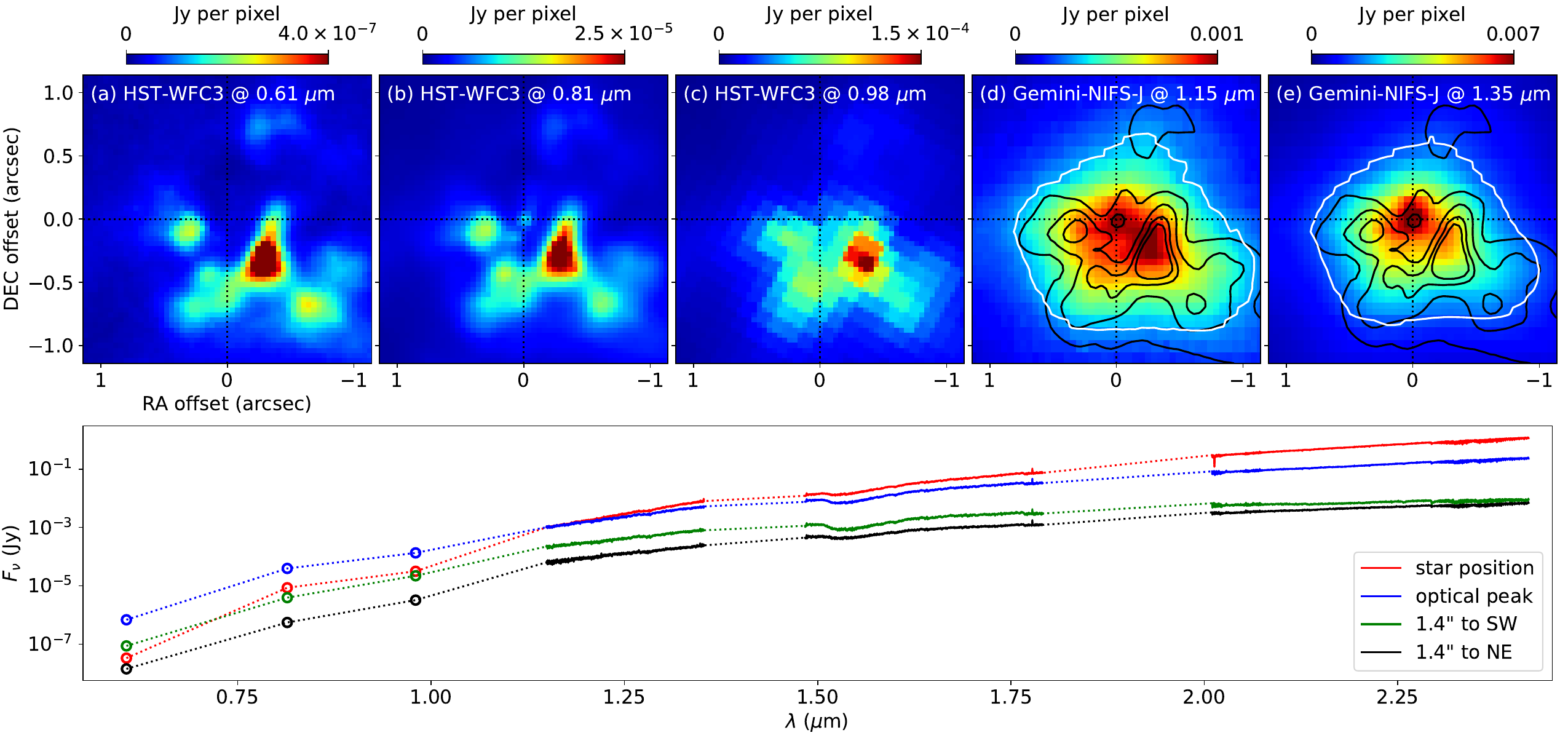}
  \caption{\label{fig:seds}
    Progressive brightening of the central stellar position along wavelength,
    with respect to the optically brightest clump located at $\sim0\farcs4$
    to the southwest.
    (a)--(c) HST images taken on 2016 May 17 with F606W, F814W, and F098M
    filters displayed in linear scales in units of Jy\ pixel$^{-1}$ with a
    regridded pixel size of $0\farcs04\times0\farcs04$ \citep[see][]{kim21}.
    (d)--(e) Gemini images in the NIFS $J$ data cube at the shortest
    wavelength of 1.15\,\micron\ and the longest wavelength of 1.35\,\micron,
    respectively, displayed in linear brightness scales.
    The white contours are at $3\times10^{-4}$\,Jy\ pixel$^{-1}$ in (d) and
    $15\times10^{-4}$\,Jy\ pixel$^{-1}$ in (e), whose angulated vertices are
    coincident with the stretched clumpy structures in black contours showing
    the brightness levels of the (b) panel at $3\times$, $6\times$, $11\times$,
    and $16\times10^{-6}$\,Jy\ pixel$^{-1}$.
    The bottom panel shows SEDs at the stellar position (red), the
    optical brightest peak (blue), and two background positions (green
    and black), derived from the HST F606W ($\sim0.606\,\micron$), F814W
    ($\sim0.814\,\micron$), and F098M ($\sim0.098\,\micron$) images, and
    Gemini $J$ (1.149--1.353\,$\micron$), $H$ (1.484--1.791\,$\micron$),
    and $K$ (2.010--2.419\,$\micron$) spectro-imaging data.
  }
\end{figure*}

The maxima of the integrated brightnesses in $J$, $H$, and $K$ bands are assumed to be located at the expected stellar positions at their observed epochs, 2016.15, 2016.16, and 2016.30, respectively. The stellar positions are calculated following the method in \citet{kim21} using the proper motion of the star CW Leo \citep{men12}.
Then, the secondary peak in $J$, progressively weakening with increasing wavelength, is well coincident with the optical brightest peak in the HST images taken at a similar epoch, 2016.38 \citep[Figure\,\ref{fig:seds}; see also][]{kim21}. In addition, the angulated vertices of the contours of the $J$-band images are coincident with the stretched clumpy structures in the optical images in position angles. The fact that these multiband images are well matched to each other approves that the maximum of each band is indeed at the expected stellar position.

The spectral energy distributions (SEDs) for four representative positions in the field of view of Gemini-NIFS observations preserve the corresponding slopes of the increasing trends in the photometric measurements from the HST images in the three filters (see the bottom panel of Figure\,\ref{fig:seds}).
The continuation in the individual SEDs again implies that the alignment between the HST and Gemini images is appropriate. Here, it is apparent that the SED at the stellar position (red curve) is much steeper than the other three lines, revealing the red nature of the carbon star.

\cite{kim21} demonstrated that a point-like source at the local brightness peak at the stellar position was identified only in the 0.8\,\micron\ image and its absence in the 0.6\,\micron\ image was explained by the red color of the carbon star. They attributed the missing of this small-scale local peak in the 1.0\,\micron\ image to flux dilution due to the 3 times lower angular resolution of such an image. We clarify this statement by applying an artificial smoothing kernel to the 0.8\,\micron\ image mimicking the low resolution of the 1.0\,\micron\ image, in which the local brightness peak at the stellar position indeed disappears. However, such inconclusive evidence for the red point-like source exactly at the stellar position in the previous work is now complemented by the progressive increase in the relative brightness at such a position along the near-infrared wavelengths. Figure\,\ref{fig:seds} clearly shows that the relative brightness of the optically brightest, southwestern region becomes progressively lower at longer wavelengths, reaching at 1.15\,\micron\ the brightness level same as that of the stellar position. It becomes much weaker than the stellar point at 1.35\,\micron\ and turns into the azimuthally averaged level of the background circumstellar envelope in the $H$ and $K$ bands.

The currently observed single-peaked distribution in the long $J$, $H$, and $K$ bands differs from the previously reported clumpy features found for quite a persistently long period (1995--2003) in near-infrared observations at high angular resolutions ($<0\farcs1$) with speckle interferometry and adaptive optics \citep{tut00,ost00,wei02,mur05,lea06}. The major clumps were arranged along a circle with the brightest clump at the southwestern part of the circle. The relative brightness of the brightest clump faded out in 2003--2005, positioning the brightest part of the central region to the rather northeastern part of the circle in various near-infrared bands \citep{wei07,ste16}. Albeit lacking astrometric information, the brightness center seems shifted to the center of the surrounding diffuse brightness distribution in the near-infrared images taken in an 2008 epoch \citep{ste16}. Our $JHK$ images in the 2016 epoch may indicate the persistence of the centralized brightness distribution since 2008, likely due to the unveiling of the carbon star; it is therefore anticipated to sustain until the stellar evolutionary phase transition, which emphasizes the necessity of a continuous monitoring of this source.

\subsection{Spatial Distribution With Respect to $r^{-2}$}\label{sec:dst}
In all $J$, $H$, and $K$ bands, the observed brightness decreases with increasing radius $r$, where $r$ corresponds to the angular distance from the predicted stellar position. Beyond a certain radius (of about 0\farcs5, depending on the band; see below), the brightness profiles are well approximated by a power law of $r$ with an index of about $-2$. Therefore, by multiplying the observed brightness in an image by $r^2$, one may find smaller-scale structures that are veiled in the bright halo distribution. This map would show a constant value at all pixels of the map if the brightness in an image indeed decreases with a power-law index of $-2$.

\begin{figure*} 
  \epsscale{1.15}
  \plotone{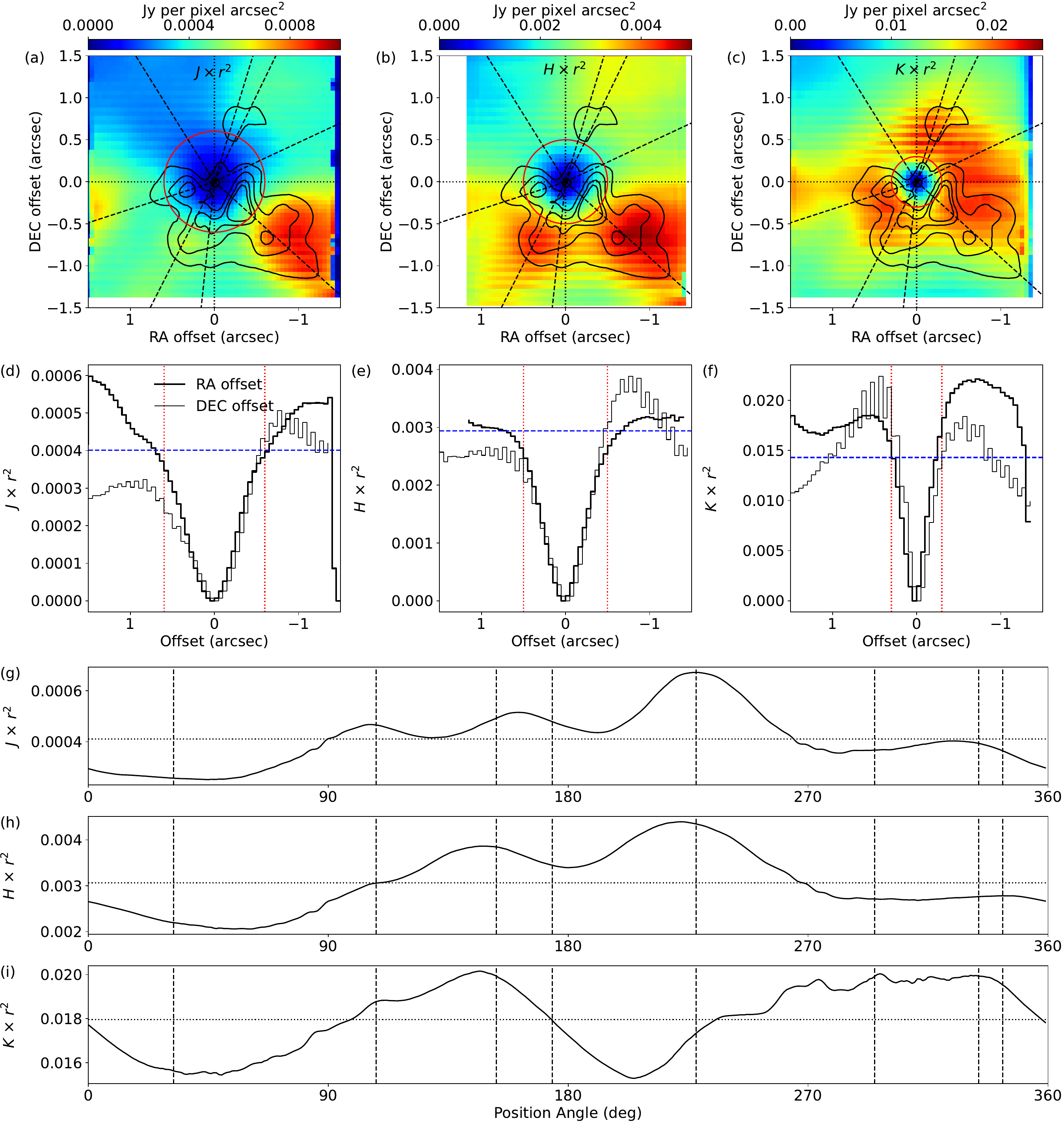}
  \caption{\label{fig:cont}
    (a)--(c) Median maps of the brightnesses of individual $J$, $H$,
    and $K$ bands along wavelength, multiplied by $r^2$ with $r$ being
    the radius around the predicted stellar position. The color scales
    are in units of Jy\,pixel$^{-1}$\,arcsec$^2$.
    For comparison, the corresponding 0.8\,\micron\ brightness distribution
    is indicated by black contours (see Figure\,\ref{fig:seds} for details).
    Black straight dashed lines correspond to the locations of searchlight
    beams identified in the HST images taken in 2011 and 2016 \citep{kim21}.
    The red circle in each panel indicates the radius at which the radial
    profile of the map changes its slope; beyond this radius, the radial
    profile either gets flattened or starts decreasing with increasing radius.
    (d)--(f) Radial profiles of the above panels along the R.A.\ and decl.\
    directions (thick and thin lines, respectively). The horizontal dashed
    line in blue indicates the median value of the map. The vertical dotted
    lines in red correspond to the radius of the red circle in the above panel.
    (g)--(i) Azimuthal profiles of the maps in (a)--(c) as a function
    of the position angle, measured from north to east. Vertical dashed
    lines indicate the position angles of the searchlight beams depicted in
    the top panels. The horizontal dotted line in each panel presents the mean
    value of the profile.
  }
\end{figure*}

Figure\,\ref{fig:cont} presents maps of individual bands (median averaged over wavelength) after being multiplied by $r^2$. These maps exhibit the position-angle dependence of brightness distributions. The three individual maps for $J$, $H$, and $K$ show brightness excesses near the optically bright, clumpy regions (marked by black contours) and along some of the searchlight beams identified by \citet{kim21} in larger $\sim5\arcsec$ scale optical images (marked by black straight dashed lines). In particular, the searchlight beams located in the southeast (at position angles of 108\arcdeg\ and 153\arcdeg), the beam in the southwest (at a position angle of 228\arcdeg), and the northern beam (position angles between 334\arcdeg\ and 343\arcdeg) well match in location with the brightest parts in Figure\,\ref{fig:cont}(a)--(c).
The southern beam, which was barely detected at a position angle of 174\arcdeg\ from the HST optical image, does not show a clear feature in this figure. The northwestern beam at a position angle of 295\arcdeg\ was the brightest beam in the 2011 HST image and was relatively weakened in the 2016 HST image, compared to the southern beams. In the near-infrared images taken in 2016, the brightness of this northwestern beam is just at the average of the image, surrounded by brighter northern and southern beams.
Along the northeastern radial beam at a position angle of 32\arcdeg\ east of north, however, all three maps do not present an excess but show a deficit. In the $J\times r^2$ map, in particular, the northeastern quadrant is significantly below the median value of the map, marked by the horizontal line in panel (d). It is also worth noting that the optical searchlight beams coincide with the locations of gaps in the anisotropic (clumpy) distributions found in molecular emission maps in the central atmospheric region \citep[see Extended Data Figures 1 and 2 of][]{vel23}. The gap coincidence despite the 3 yr epoch difference is consistent with the stable locations of the searchlight beams observed in the optical over 15 yr \citep{kim21}.

The position-angle dependence in excess or deficit over the $r^{-2}$ distribution is significant in $J$, and rather moderate in $H$ and $K$. In order to measure the azimuthal variation in each image, a profile as a function of position angle is derived by averaging the values within an annulus of width of 0\farcs5 with its inner edge at the red circle drawn in the top panels of Figure\,\ref{fig:cont}. The maximum to minimum ratios of the resulting profiles plotted in (g)--(i) are 2.7, 2.1, and 1.3, respectively. Such a wavelength dependence that the variation is larger at shorter wavelengths, suggests that its origin is at the scattering of stellar light due to circumstellar dust grains.

The central part cannot be flat in these maps, because the $r^{-2}$ profile yields a mathematical singularity at the center. Observational seeing effects lower the values within the central 0\farcs1--0\farcs2, or slightly more extended region. In addition, as a physical reason, dust grains start to form at a certain radius, $\sim2$--9 stellar radii depending on the grain species, where the temperature of the material ejected from the star drops below the typical dust condensation temperatures of 1000--1500\,K \citep{hof18}. Within such a radius ($\sim0\farcs14$ for CW Leo), therefore, the wind velocity has not reached the terminal velocity and the density profile of the wind does not follow the same power-law slope of $-2$ as that of the outer part of the circumstellar envelope. The wind acceleration curve of CW Leo, empirically derived by \citet{dec15}, confirms that the wind velocity is nearly constant beyond $r\sim0\farcs2$.
In Figure\,\ref{fig:cont}(d)--(f), the radii marked by vertical dotted lines in red (0\farcs6, 0\farcs5, and 0\farcs3 in $J$, $H$, and $K$, respectively) are the radii at which we identify the significant slope changes in the radial profiles. Beyond these radii, the radial profiles either become flattened or decrease with radius, implying that the brightness falls as $r^{-2}$ or slightly steeper than that.

\subsection{Color maps}\label{sec:clr}
We have constructed three color maps using the median flux densities in the individual bands and assuming the Vega system\footnote{https://www.gemini.edu/observing/resources/magnitudes-and-fluxes \citep[see also][]{tok05,hew06}.}.
As displayed in Figure\,\ref{fig:cmap}, the $(H-K)$ and $(J-K)$ maps show redder colors toward the stellar position. These color distributions are somewhat asymmetric; a relatively bluer color to the southwestern direction is clearly presented, in particular in the $(J-K)$ map. This is consistent with the fact that most of southwestern quadrant of the map is overlaid by the optically bright, thus blue, features.

On the other hand, the $(J-H)$ map presents very different distribution. Redder $(J-H)$ colors are present along the northern direction starting from the stellar position. This color distribution is analogous with the color distribution in the optical wavelengths, measured by \citet{kim21} using the brightness ratio between the images at 0.6 and 0.8\,\micron. The redder color toward the north and northeast in the $(J-H)$ color map attributes to the relative lack of $J$ brightness in the northeast quadrant (see Figure\,\ref{fig:cont}(a)).

\section{ABSORPTION FEATURES}\label{sec:spe}
\subsection{Spectroscopic Indices}\label{sec:ind}
\begin{figure*} 
  \epsscale{1.15}
  \plotone{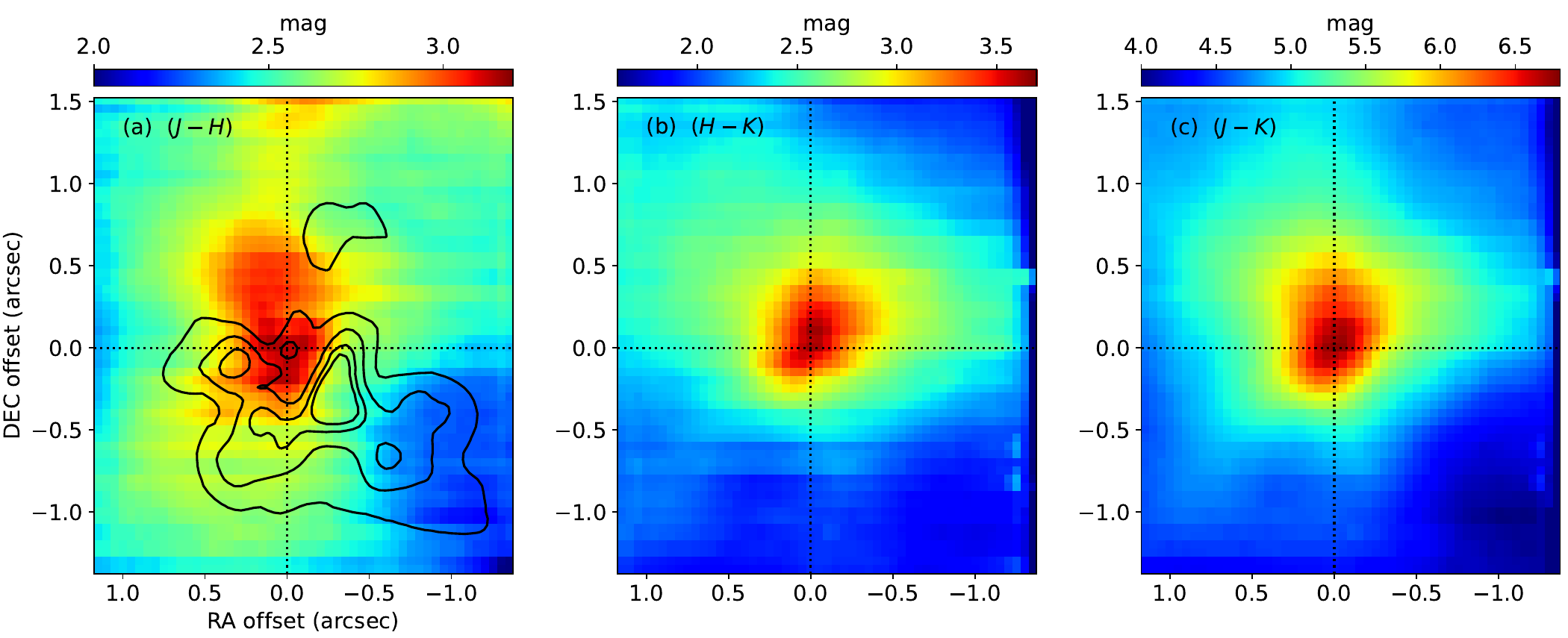}
  \caption{\label{fig:cmap}
    Three combinations for magnitude difference between the $JHK$-band
    images: (a) $(J-H)$, (b) $(H-K)$, and (b) $(J-K)$ color maps. The
    redder color of the maps indicates the intrinsically red nature of
    the central star. Black contours display the corresponding 0.8\,\micron\
    brightness distribution (see Figure\,\ref{fig:seds} for details).
  }
\end{figure*}

CW Leo is extremely red; the $(J-K)$ color at the stellar position is $>6$ and the central circumstellar envelope at $r\la1\arcsec$ is well above 4 (see Figure\,\ref{fig:cmap}(c)). For comparison, the synthetic color of $(J-K)\la1.6$ is the limit of hydrostatic dust-free models in reproducing the spectroscopic and photometric properties of carbon stars, and additional reddening or circumstellar dust emission is invoked for redder stars \citep[e.g.,][]{ari09,gon17}. The red nature of CW Leo is thus consistent with its high mass-loss rate of $>2\times10^{-5}\,\Mspy$ \citep[e.g.,][]{gue18,fon22}.

\begin{figure*} 
  \epsscale{1.15}
  \plotone{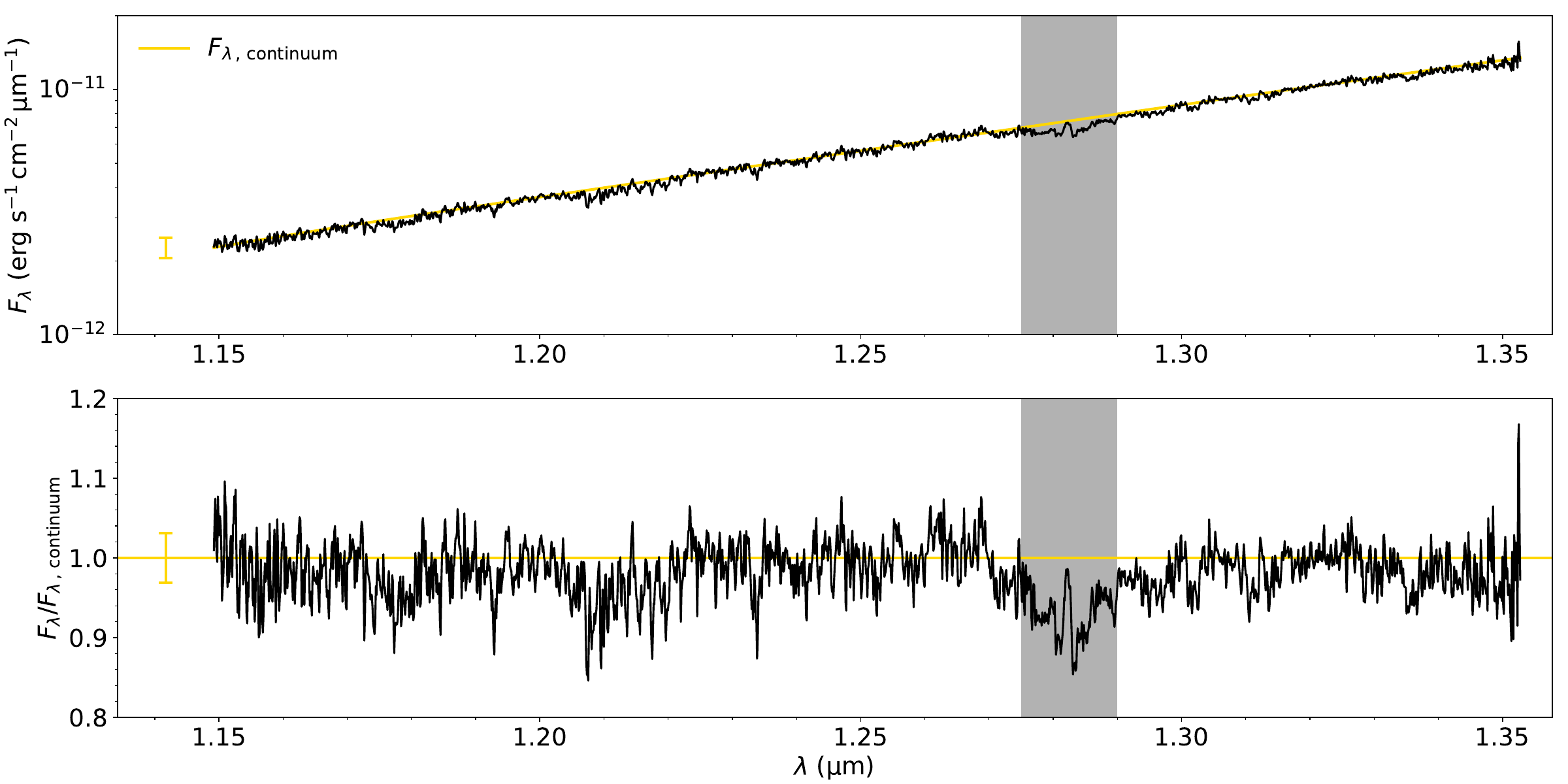}
  \caption{\label{fig:Jspc}
    Top: $J$-band spectra at the central pixel in the Gemini-NIFS
    spectro-imaging data (black). A least-squares third-degree polynomial
    fits used to define the continuum flux density (yellow).
    The area masked in gray is not used in the continuum fitting.
    In the 1.275--1.290\,\micron\ range, an artifact remains after
    telluric correction. The yellow bar plotted on the bottom left
    shows the residual standard error of the fit, mostly attributed
    to weaker spectral features that are not considered in this paper.
    Bottom: spectrum normalized by the continuum as a result
    of the polynomial fit. The yellow bar on the left presents the
    uncertainty propagated from continuum fitting.
  }
\end{figure*}

\begin{figure*} 
  \epsscale{1.15}
  \plotone{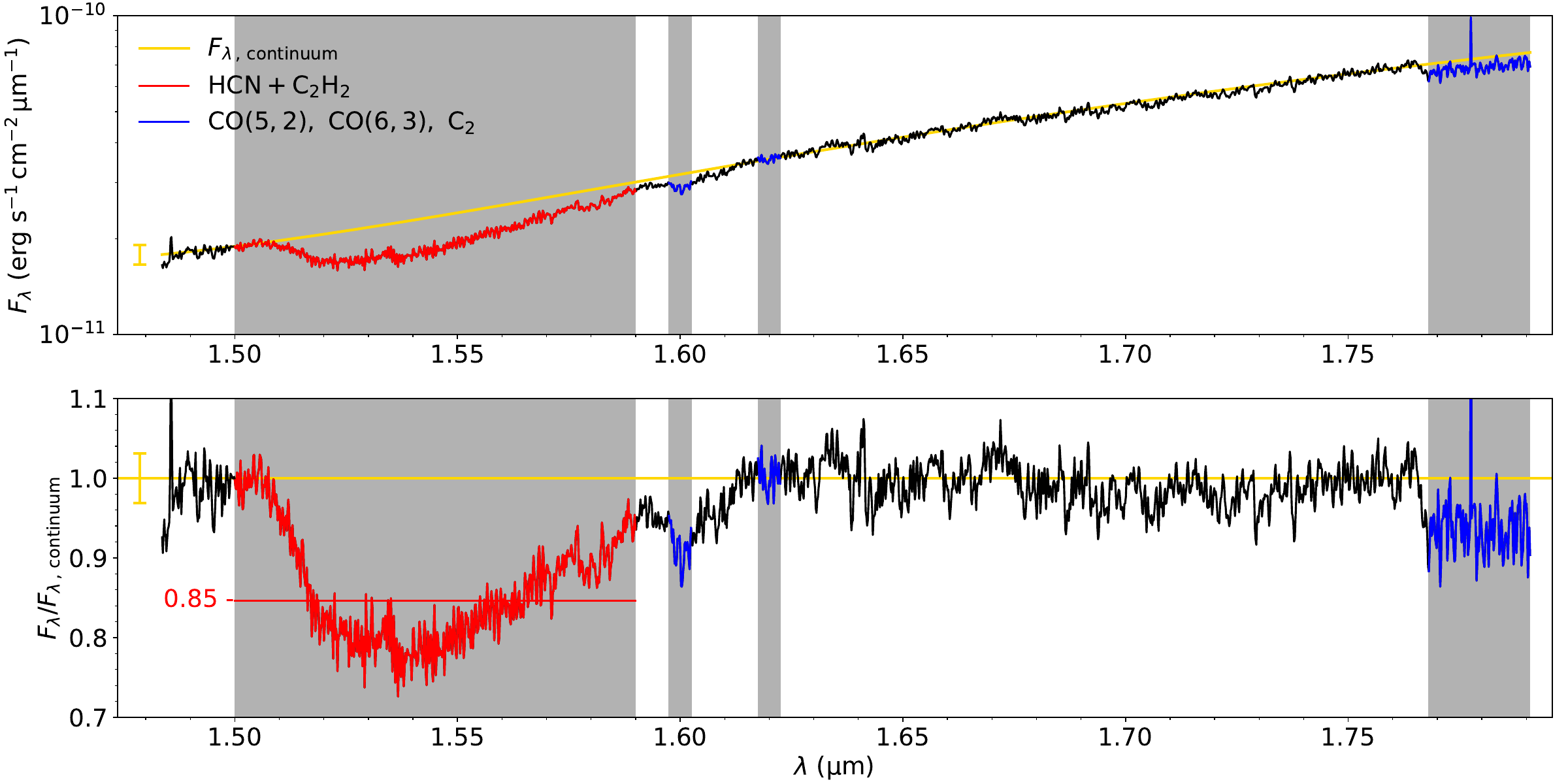}
  \caption{\label{fig:Hspc}
    Same as Figure\,\ref{fig:Jspc} but for $H$-band spectra at the central
    pixel.
    The 1.53\,\micron\ absorption feature (DIP153; 1.50--1.59\,\micron),
    marked in red, is prominent. The median of the normalized spectrum
    is 0.85, corresponding to a spectroscopic index of 0.18 (see text).
    The COH$_{52}$ (1.5974--1.6026\,\micron), COH$_{63}$ (1.6174--1.6226%
    \,\micron), and C2 (1.7680--1.7909\,\micron) absorption features are
    marked in blue.
  }
\end{figure*}

\begin{figure*} 
  \epsscale{1.15}
  \plotone{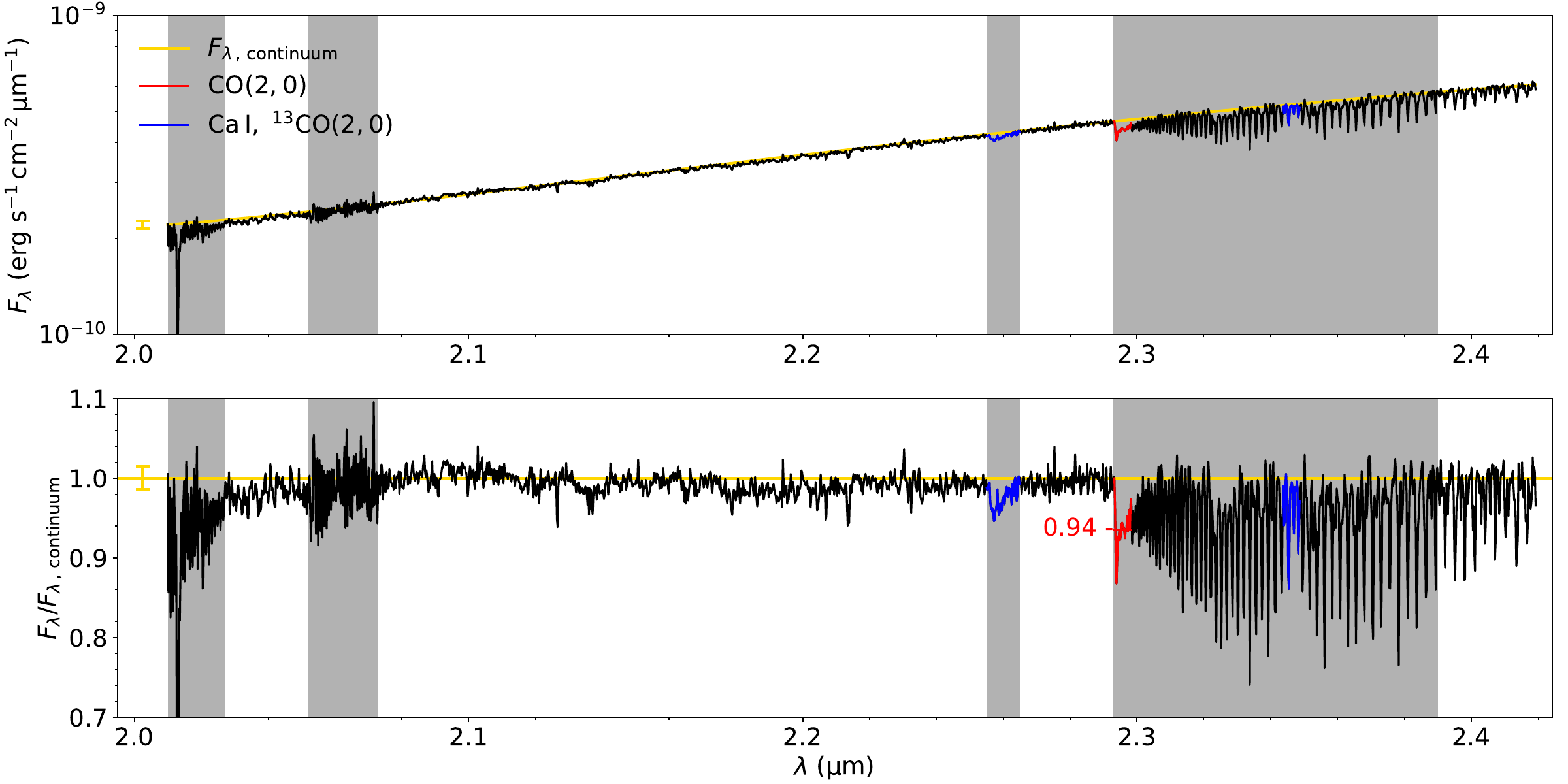}
  \caption{\label{fig:Kspc}
    Same as Figure\,\ref{fig:Jspc} but for $K$-band spectra at the central
    pixel. The CO12 (2.2931--2.2983\,\micron) absorption feature is marked
    in red, and the median value of its normalized spectrum is indicated as
    0.94. The CaI (2.255--2.265\,\micron) and CO13 (2.3436--2.3488\,\micron)
    absorption features are marked in blue.
  }
\end{figure*}

All of the $J$-, $H$-, and $K$-band continuum spectra at the stellar position of CW Leo are well defined empirically by $F_\lambda\propto\lambda^\alpha$ with a spectral index $\alpha$ of $10.9\pm0.02$, $8.1\pm0.02$, and $5.7\pm0.01$, respectively. The uncertainties in the slope measurements mainly indicate spectral fluctuations due to unresolved features. The large slopes of SEDs in the individual bands can be seen in Figures\,\ref{fig:Jspc}--\ref{fig:Kspc}.
In these figures, however, instead of the above power-law fitting, we define the continuum spectral distribution with a least-squares fit to a polynomial of degree 3 (yellow line); we compared these with carbon star templates over a range of $(J-K)$, which do not generally present single power-law distributions in the near-infrared \citep[see][]{gon16}.

Among the characteristic absorption features of the carbon star templates whose spectroscopic indices were measured by \citet{gon16}, seven features are identified in the $H$ and $K$ spectra of CW Leo and their wavelength ranges are as given in Table\,\ref{tab:list}. The broad and deep 1.53\,\micron\ feature (dubbed as DIP153) is particularly outstanding, as shown in Figure\,\ref{fig:Hspc}. The 1.53\,\micron\ absorption feature was first identified in a carbon-rich Mira-class variable star as an overtone of the already-known strong absorption band at 3.1\,\micron\ dominated by HCN and C$_2$H$_2$ molecules \citep{goe81}.

With the above fitting results for the continuum, $F_{\lambda,\rm\,continuum}$, we measure the strength of an absorption feature $X$ using the formula,
\begin{equation}\label{eqn:indx}
  I(X) = -2.5 \log_{10} <F_\lambda(X)/F_{\lambda,\rm\,continuum}(X)>,
\end{equation}
where the normalized flux density is averaged over the wavelength bin for absorption feature $X$ in a median base. We measure these spectroscopic indices at the center of the NIFS images (red in Figure\,\ref{fig:indx}), an offset position at 1\farcs2 to the north (blue), and integrated over the field of view of the images (green).

\begin{deluxetable}{cccc}
  \tablecaption{\label{tab:list}%
    Properties of Our spectroscopic indices.}

  \tablehead{\colhead{Index Name} & \colhead{Relevant Elements}
    & \colhead{$\lambda_{\rm min}$ (\micron)}
    & \colhead{$\lambda_{\rm max}$ (\micron)}}

  \startdata
  \tableline
  DIP153      & HCN\,$+$\,C$_2$H$_2$ & 1.5000 & {\bf 1.5900}\\
  COH$_{52}$  & CO$(5,2)$        & 1.5974 & 1.6026\\
  COH$_{63}$  & CO$(6,3)$        & 1.6174 & 1.6226\\
  C2          & C$_2$          & 1.7680 & {\bf 1.7909}\\
  \tableline
  CaI         & \ion{Ca}{1}    & {\bf 2.2550} & {\bf 2.2650}\\
  CO12        & CO$(2,0)$        & 2.2931 & 2.2983\\
  CO13        & $^{13}$CO$(2,0)$ & 2.3436 & 2.3488\\
  \enddata

  \tablecomments{The wavelength ranges of absorption bands are adopted
    from \citet{gon16}, except for the values in bold, which are newly
    defined in this paper. The index for COH is defined following
    \citet{gon16} as the mean of the COH$_{52}$ and COH$_{63}$ indices:
    $\rm COH=(COH_{52}+COH_{63})/2$.}
\end{deluxetable}

These new definitions for spectroscopic indices are conceptually similar to the ones adopted in earlier works \citep[e.g.,][]{wor94,gon16}, at which, however, the (pseudo-)continuum was chosen as the mean flux density in a predefined wavelength bin adjacent to the molecular band of interest. When we attempt to measure the spectroscopic indices following the previous definitions, some of the absorption features lie above the levels of the pseudocontinuum, making false emission features with the calculated indices as being negative values (see Appendix\,\ref{sec:gon}). This problem is serious for redder carbon stars with extremely steep SED slopes in the near-infrared, like CW Leo. Therefore, we have newly defined the continuum by a global fitting rather than adopting an adjacent value. We also apply the new definitions of spectroscopic indices to the spectra of a sample of carbon stars provided by \citet{gon16}, for comparison, as displayed in Figure\,\ref{fig:indx}.

The DIP153, COH, CO12, and CO13 indices of the template stars show overall decreasing trends as a function of $(J-K)$; these trends are particularly strong for stars with relatively deep DIP153 features (filled black circles). The C2 index of the template stars shows an increasing trend along $(J-K)$ for the bluer stars with weaker DIP153 features (open circles), while it is rather flat (or slightly decreasing) for stars with strong DIP153 features (filled circles). In addition, a small but definite absorption feature found at 2.255--2.265\,\micron\ in Figure\,\ref{fig:Kspc} was not defined in \citet{gon16}, where we newly measure the CaI index. This spectroscopic index tends to have a maximum at $(J-K)$ of a moderate value $\sim2.5$.

The DIP153 feature is mainly attributed to a combination of C$_2$H$_2$ and HCN and dominantly appears at the minimum phase of dynamical models of sufficiently cool and dense winds \citep[see Figure 14 of][]{gau04}. Under such a cool condition, C$_2$H$_2$ forms at the chemical expense of C$_2$ \citep{gro09}, which may explain the anticorrelation of the trends between these two molecular DIP153 and C2 features of the template carbon stars of \citet{gon16} at $(J-K)\la3$ (see Figure\,\ref{fig:indx}). The weakening of C2 and CaI features for the redder carbon stars with $(J-K)\ga3$ would be explained, on the other hand, by a dilution effect whereby the additional thermal continuum emission by circumstellar dust overspreads the absorption trough in the spectrum, as suggested by \citet{gon16}.

The spectroscopic indices of CW Leo (plotted in red, blue, and green colors) are larger than the trends (magenta) found from the template carbon stars having strong DIP153 features, like CW Leo. In particular, the DIP153 index of CW Leo is significantly larger than in any other template stars.
Indicated by the very large slope of SED of CW Leo, compared to those of the template carbon stars, the dilution of absorption features due to the dust thermal continuum emission may be tremendous. The nevertheless extremely strong 1.53\,\micron\ feature may imply its circumstellar origin at sufficiently large distances from the carbon star, where the circumstellar material cools down.
It corresponds to the richness of circumstellar molecules, which is expected to be proportional to the high mass-loss rate of CW Leo and consistent with the large $(J-K)$ color indicating additional reddening or thermal emission due to circumstellar dust.

\subsection{Spatial Distributions of Absorption Features}\label{sec:ssd}
Figure\,\ref{fig:imap} displays the spectroscopic indices showing meaningful variation in the distribution over the observed images. The absorption depths of COH$_{63}$, C2, CaI, and CO13 are, on average over the field of view, less than twice the spectral fluctuations due to the unresolved spectral features in the $H$ and $K$ bands. On the other hand, the DIP153, CO12, and COH$_{52}$ absorption features are distinguishable from the unresolved spectral fluctuations by factors of 5.9, 4.5, and 2.9 on average, respectively (see Figure\,\ref{fig:imap}).

Interestingly, the DIP153, COH$_{52}$, and CO12 indices are relatively smaller to the north--northeast. In addition, the deficit regions are in a cone shape with an opening angle that seems to be restricted by the optically brightest structures enshrouding the star.
Recalling the bipolar cavity scenario of, e.g., \citet{men01}, these northern regions with less absorption features may indicate a deficit of relevant molecules within the bipolar cavity cones, which include less gaseous material. In this viewpoint,
albeit less significant than the northern part, the southern region of Figure\,\ref{fig:imap} also lacks absorption features, relative to the surrounding southwestern and southeastern regions. Also notice that the absorption-less regions coincide with the (scattering) brightness-less regions in Figures\,\ref{fig:cont}(a)--(c).

Another interesting thing, found in Figure\,\ref{fig:imap}(c), is that the radial profiles of the CO12 index map for all directions are minimized at around $r\sim0\farcs2$. This corresponds to the radius that the wind undergoes acceleration \citep{dec15}, occurring with dust formation at such a radius.
The low energy level for the CO first overtone band head occurs at $>3000$\,K \citep{bie02}, thus the CO12 absorption feature cannot originate from circumstellar molecules. Stellar light may be absorbed by CO molecules in the high-temperature atmospheric region and its reflection by dust occurs in the circumstellar region, yielding the spectral distribution that is finally observed. In this process, the addition of continuum due to dust thermal emission could reduce the depth of the absorption feature.
Dust thermal emission will be enhanced at around the radius that dust grains efficiently form and are heated by stellar radiation directly; therefore the minimization of the CO12 index map reconfirms the dust condensation radius of $\la0\farcs2$ in the case of CW Leo.

\subsection{Expansion Velocity}\label{sec:exp}
In the wavelength range of 2.293--2.420\,\micron, with the upper limit given by the boundary of the observed $K$ band, a forest of CO overtone bands is presented. A measurement using the forest of lines as a whole significantly improves the velocity measurement precision far beyond the spectral resolution \citep[e.g.,][]{ton79,all07}, thus it hints at the net velocity of dust grains that have participated in the scattering of the stellar light after penetrating the CO absorption zone. A cross correlation of the spectrum at an arbitrary position with that of the central position yields a wavelength lag between two spectra, which is converted to a velocity lag with respect to the center. By applying this to all pixels in the $K$-band image, a map of velocity lag of dust grains acting as ``moving mirrors'' for stellar light is achieved (Figure\,\ref{fig:vexp}). Positive velocities are obtained throughout the map, indicating the outward motion of dust grains from the star position at the center \citep[see, e.g.,][]{yos11}.

The velocities measured with this method at all pixels of the image are profiled as a function of radius, and the profile is compared to the wind expansion velocity curves of CW Leo suggested based on the maximum Doppler velocities and sizes of molecular line emissions in various transitions \citep{fon08,dec15}. Figure\,\ref{fig:vexp}(b) shows that our measurement via CO overtone bands yields a smaller velocity in the central 1\arcsec\ region than the actual expansion velocity at a given radius.
Since the velocity measured here is the integration of the expansion velocities of the dust grains that the photon has met during the entire scattering process, this result may indicate that scattering occurs multiple times in the wind acceleration zone, significantly lowering the net velocity.
On the other hand, in the measurement at a larger radius, the portion of fully accelerated dust grains that were encountered during scattering would be larger and the measured velocity reaches the terminal velocity of the wind.

The measured velocity map also presents a position-angle dependence. Figure\,\ref{fig:vexp}(c) displays that south--southwestern (180\arcdeg--270\arcdeg) velocities are faster than in the northern part.
Notice that \citet{kim21} measured the transverse velocity of the wind using differential proper motion of the circumstellar whorled pattern in two optical images with an epoch difference of 5\,yr, reporting a systematical reduction of the expansion velocity to the north. It was proposed that this result hints at an eccentric-orbit binary viewed nearly face on and having the pericenter of the carbon star projected onto the northern direction, through its comparison with the velocity measurements of hydrodynamic models (\citealp{kim23}; see also Figure\,4 of \citealp{kim23iau}).
Despite the difference in the exact elements of the measured velocities, the position-angle dependence of the wind expansion velocity is consistent in the two different methods.

\section{SUMMARY AND DISCUSSION}\label{sec:dis}
We have observed the central region of the famous carbon star CW Leo in the near-infrared $J$, $H$, and $K$ bands using NIFS equipped on the Gemini-North telescope. The observation epochs from 2016 February to April are very close to the epoch of optical imaging observations using the HST on 2016 May 17, facilitating astrometry despite the small field of view of NIFS. The major findings in this paper are listed below.

\begin{itemize}
\item At the 2016 epoch, the near-infrared $H$- and $K$-band data cubes exhibit single-peaked brightness distributions in all wavelengths, differing from the epochs of 1995--2003 when several clumps were presented.
\item The $J$-band data cube reveal two brightest regions, among which the brighter region's relative brightness progressively increases along wavelength, which is located at the single, brightest central position of the $H$ and $K$ bands.
\item The second peak in the $J$-band data cube becomes brighter, relative to the first peak, as the wavelength decreases. By locating this second peak to the optically brightest position in the HST optical images, the primary brightness peak in the $JHK$ bands nicely coincides with the stellar position in the sky coordinates derived from the astrometry using the field stars in the HST images.

\item Small variation in brightness, unveiled through multiplication by $r^2$, presents the position-angle dependence corresponding to the radial beam pattern found in the optical images. Light deficit to the northeast direction is distinct and occurs across all the near-infrared wavelengths.

\item The redder $(J-H)$ color lies in the northern direction, while the other two color combinations are redder toward the stellar position.

\item CW Leo's $(J-K)$ color is extremely red: $>6$ at the central pixel of the NIFS images, and $>4$ for the central 1\arcsec\ region in radius.
\item The continuum spectra of CW Leo in the near-infrared can be approximated as a power-law SED with a power-law index of 11, 8, and 6 for $J$, $H$, and $K$, respectively.
\item Spectroscopic indices of six absorption features are derived using an updated method. Compared to the template carbon stars in the literature, CW Leo's absorption features are surprisingly deeper than the predictions of the atmospheric models. The strong absorption features may prove the active formation of circumstellar molecules, which is consistent with the high mass-loss rate and the large $(J-K)$ color indicating additional thermal emission or reddening due to circumstellar dust.

\item Spatial distributions of the absorption features DIP153, COH$_{52}$, and CO12 clearly show a deficit to the north--northeast (and probably also to the south--southwest), supporting a (bipolar) cavity cone scenario.
\item CO12 absorption minimum at $r\sim0\farcs2$ corresponds to the dust formation radius, where the veiling of absorption features due to dust thermal emission would be maximized.

\item Velocity measured through the cross correlation of CO overtone bands, scattered by dust, shows a slower increase with radius than the acceleration curve of molecular gas; it reaches the terminal velocity at the radius beyond 1\arcsec. On the other hand, the measured velocity has a position-angle dependence, in which the velocity to the north is minimized.
\end{itemize}

\acknowledgments
We thank the anonymous referee for helpful suggestions that improved this paper. We are grateful for the valuable comments provided by Dr.\ Andreea Petric, Dr.\ Kwang-Il Seon, and Dr.\ Seok-Ho Lee.
This research was supported by the National Research Foundation of Korea (NRF) grant (No.\ 2021R1A2C1008928) and the Korea Astronomy and Space Science Institute (KASI) grant (Project No.\ 2023-1-840-00), both funded by the Korean government (MSIT). 
Y.O.\ acknowledges the support by the National Science and Technology Council (NSTC) of Taiwan through grant 2112-M-001-021-.
J.H.K.\ acknowledges the support from the NRF grant, No.\ 2021M3F7A1084525.
This work is based on observations supported by K-GMT Science Program (Program ID: GN-2016A-Q-18) of KASI and obtained at the international Gemini Observatory, a program of NSF NOIRLab, which is managed by the Association of Universities for Research in Astronomy (AURA) under a cooperative agreement with the U.S. National Science Foundation on behalf of the Gemini Observatory partnership: the U.S. National Science Foundation (United States), National Research Council (Canada), Agencia Nacional de Investigaci\'{o}n y Desarrollo (Chile), Ministerio de Ciencia, Tecnolog\'{i}a e Innovaci\'{o}n (Argentina), Minist\'{e}rio da Ci\^{e}ncia, Tecnologia, Inova\c{c}\~{o}es e Comunica\c{c}\~{o}es (Brazil), and Korea Astronomy and Space Science Institute (Republic of Korea). This work was enabled by observations made from the Gemini-North telescope, located within the Maunakea Science Reserve and adjacent to the summit of Maunakea. We are grateful for the privilege of observing the Universe from a place that is unique in both its astronomical quality and its cultural significance.

\begin{figure*} 
  \epsscale{1.15}
  \plotone{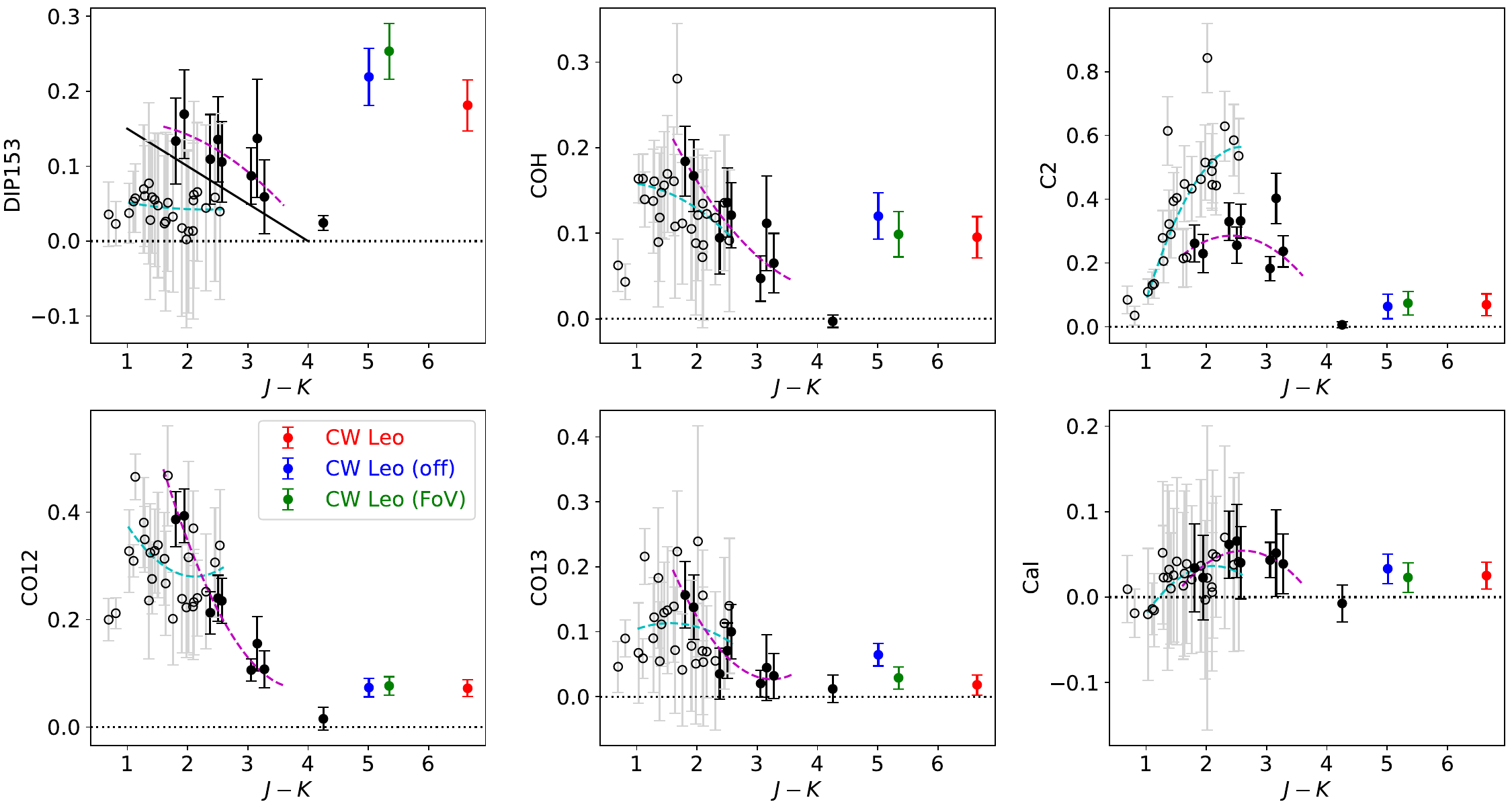}
  \caption{\label{fig:indx}
    Spectroscopic indices derived from our Gemini-NIFS spectro-imaging data
    on the central region of circumstellar envelope of CW Leo: the stellar
    position (red), an offset position (blue; 1\farcs2 to the north), and
    the integration over the field of view (green). The corresponding indices
    of a sample of carbon stars are marked by black filled and open circles,
    which are separated by the black solid line drawn in the top left panel
    for the DIP153 indices \citep[see][]{gon16}. Vertical bars indicate
    uncertainties propagated from the residual standard errors of the fits
    for individual continuum determinations.
    Magenta and cyan dashed lines represent polynomial fits of degree
    two for the template carbon stars marked by filled and open circles,
    respectively, excluding three outliers at $(J-K)<1$ and $>4$. The star
    V CrA at $(J-K)=4.3$, which was excluded during the fit, still follows
    the fitting results (magenta dashed lines).
  }
\end{figure*}

\begin{figure*} 
  \epsscale{1.15}
  \plotone{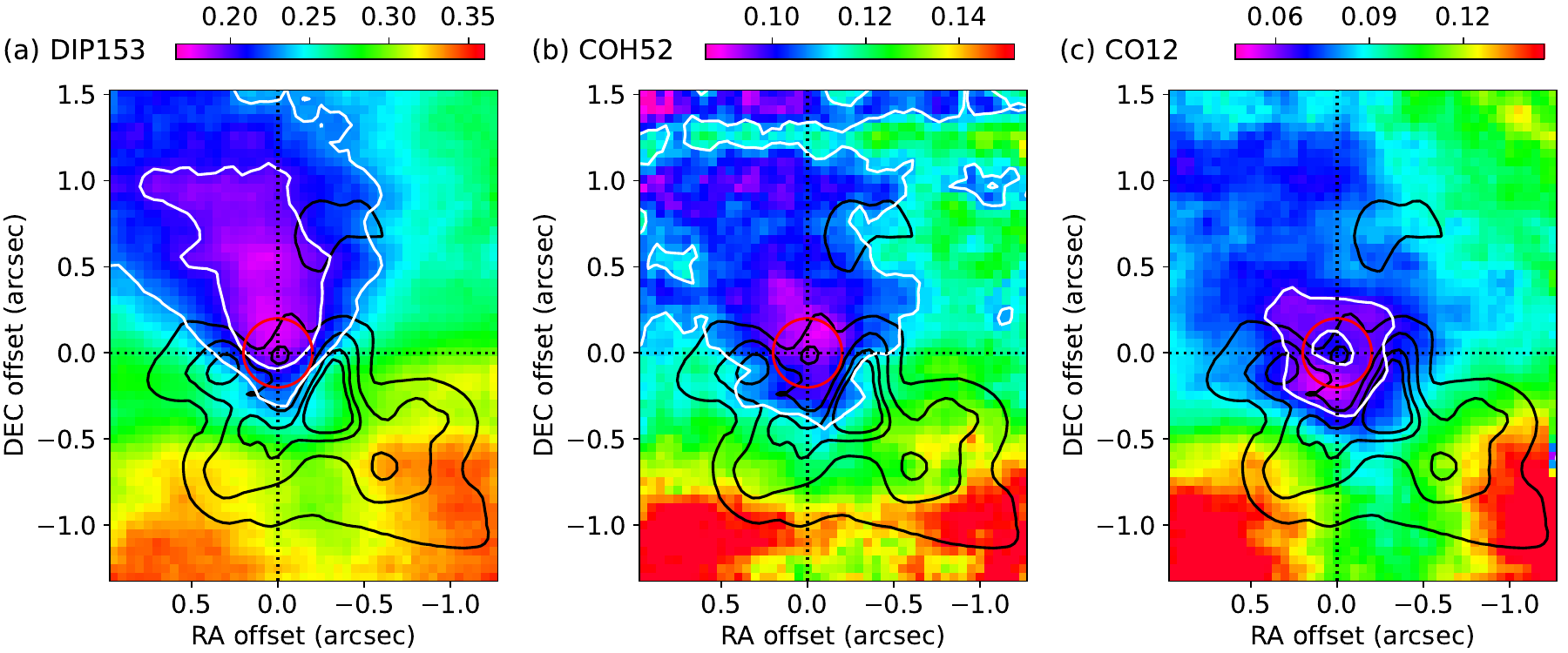}
  \caption{\label{fig:imap}
    Maps of the spectroscopic indices (a) DIP153, (b) COH$_{52}$, and (c) CO12,
    showing meaningful variation within the scale of the observed image.
    The corresponding 0.8\,\micron\ brightness distribution is marked by
    black contours, same as in Figures\,\ref{fig:seds} and \ref{fig:cmap}.
    Red circle indicates a radius of 0\farcs2 from the center, at which
    wind acceleration occurs \citep{dec15}. White contours show the levels
    of 0.20 and 0.23 of the spectroscopic index map in the (a) panel, 0.11
    in the (b) panel, and 0.065 in the (c) panel. Color bars are labeled from
    $-2\sigma$ to $2\sigma$ about the mean value of the maps, where $\sigma$
    indicates the standard deviation of the map.
  }
\end{figure*}

\begin{figure*} 
  \epsscale{1.15}
  \plotone{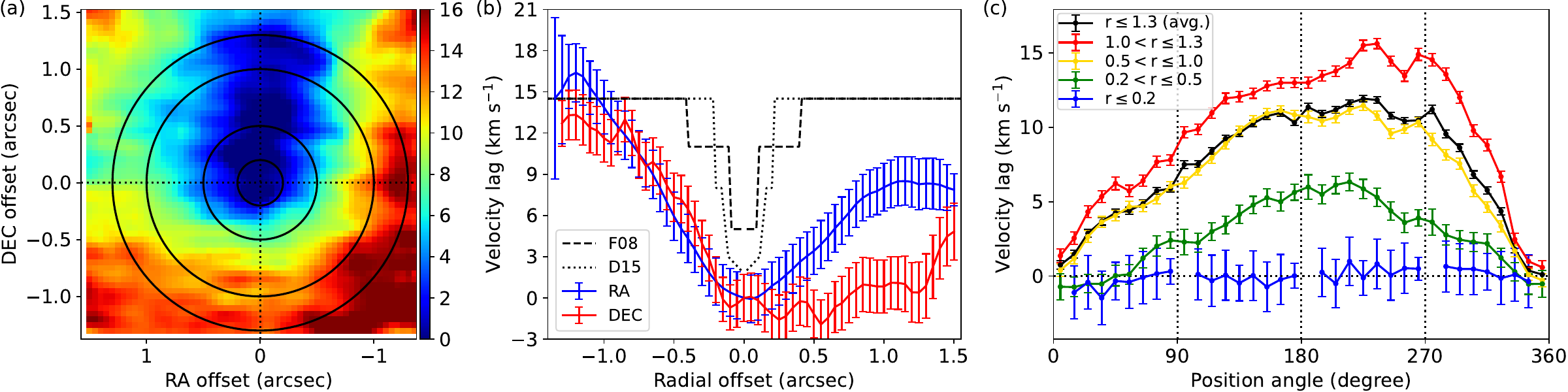}
  \caption{\label{fig:vexp}
    Velocity lag derived from cross correlations of CO overtone bands
    observed in the range of 2.293--2.420\,\micron.
    (a) The measured velocity map in a linear color scale from 0 to
    16\,\kmps\ shows an overall increase with radius from the expected
    stellar position, but also displays anisotropy. The radii of the black
    circles are 0\farcs2, 0\farcs5, 1\farcs0, and 1\farcs3, respectively.
    (b) The profiles along the west--east (blue) and south--north (red)
    cuts, including the stellar position at the center, are obtained
    from the (a) panel. The expansion velocity curves of the wind of CW Leo
    suggested by \citet[F08]{fon08} and \citet[D15]{dec15} are plotted
    by dashed and dotted lines, for comparison.
    (c) The velocity lag profiles as a function of position angle are
    measured at different radii.
  }
\end{figure*}

\clearpage
\appendix
\section{Spectroscopic Indices as Defined in Gonneau et al.\ (2016)}
\label{sec:gon}
\begin{figure*}[!b] 
  \epsscale{1.15}
  \plotone{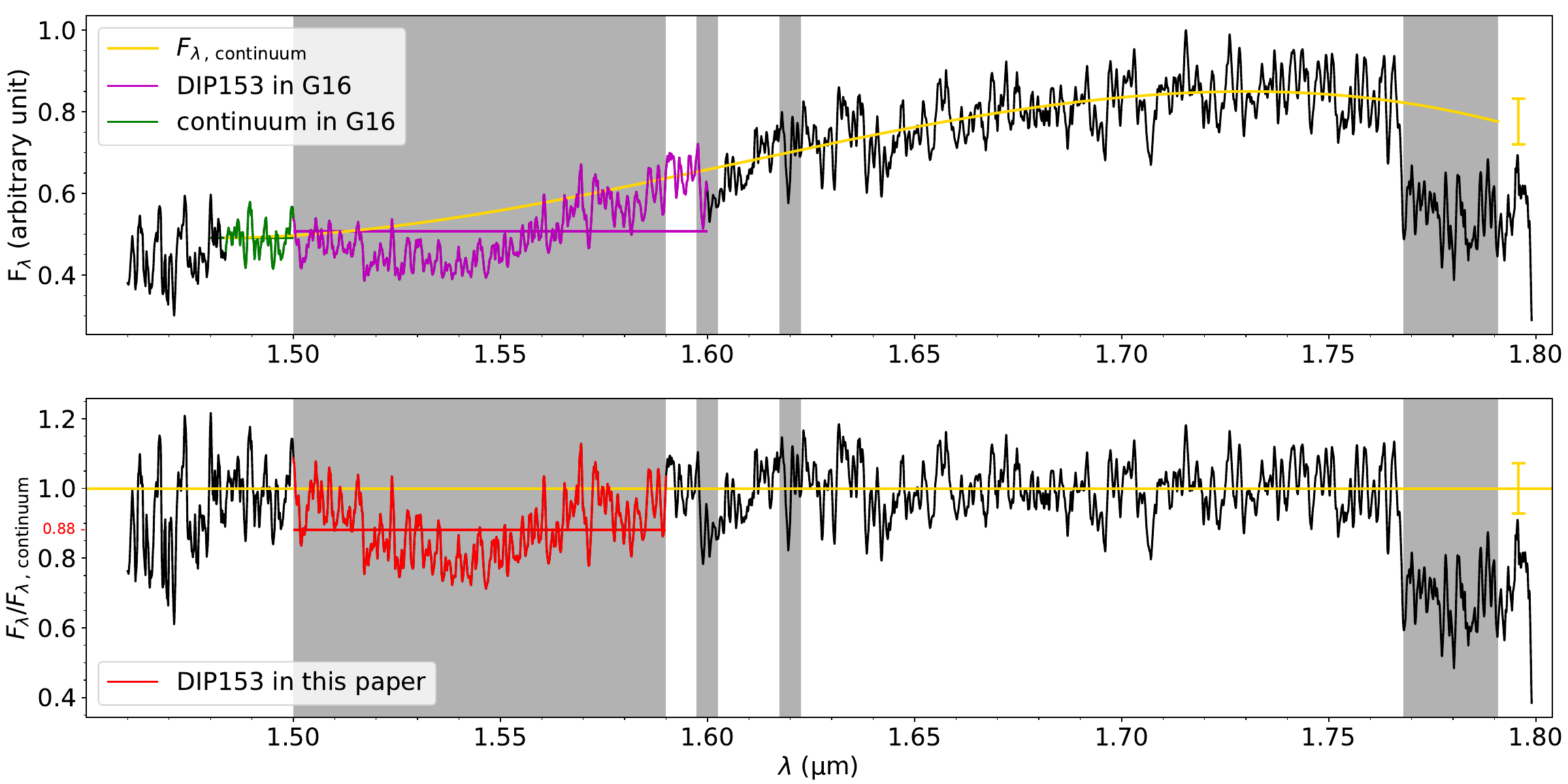}
  \caption{\label{fig:exam}
    Comparison between the spectroscopic indices based on the measurement
    methods in this paper and in \citet[G16]{gon16}. The $H$-band spectrum
    of the carbon star SHV 0536139-701604 (black) is used as an example.
    The spectral ranges for the DIP153 feature and its pseudocontinuum
    defined in G16 are marked by magenta and green colors in the upper
    panel. Their mean values, as indicated by the horizontal lines, are
    inserted to Equation\,(\ref{eqn:indx}), yielding a spectroscopic
    index of $-0.04$, literally indicating emission.
    In this paper, the global continuum curve (yellow in the upper panel)
    is used to determine the normalized spectrum (lower panel), from which
    the median value of the spectral range denoted by red color results in
    a spectroscopic index of 0.14; the positive index properly represents
    an absorption feature. See Figure\,\ref{fig:Hspc} for details.
  }
\end{figure*}

The spectroscopic indices of the absorption features derived in this paper are conceptually similar to those in \citet{gon16} and reference therein, except for the different definition of the continuum as the base of the line. Figure\,\ref{fig:exam} demonstrates the detailed difference in the two methods of measurement of spectroscopic index, for example, of the DIP153 feature for an example carbon star, SHV 0536139-701604. For a star having a large slope in its SED, an absorption feature, especially a broad feature, may have larger fluxes at the longest ranges of the band pass than the flux in the continuum band pass that was defined to be adjacent shorter wavelengths. The false emission effects owing to this become more serious for stars with extremely large SED slopes, like CW Leo.

\begin{figure*} 
  \epsscale{1.15}
  \plotone{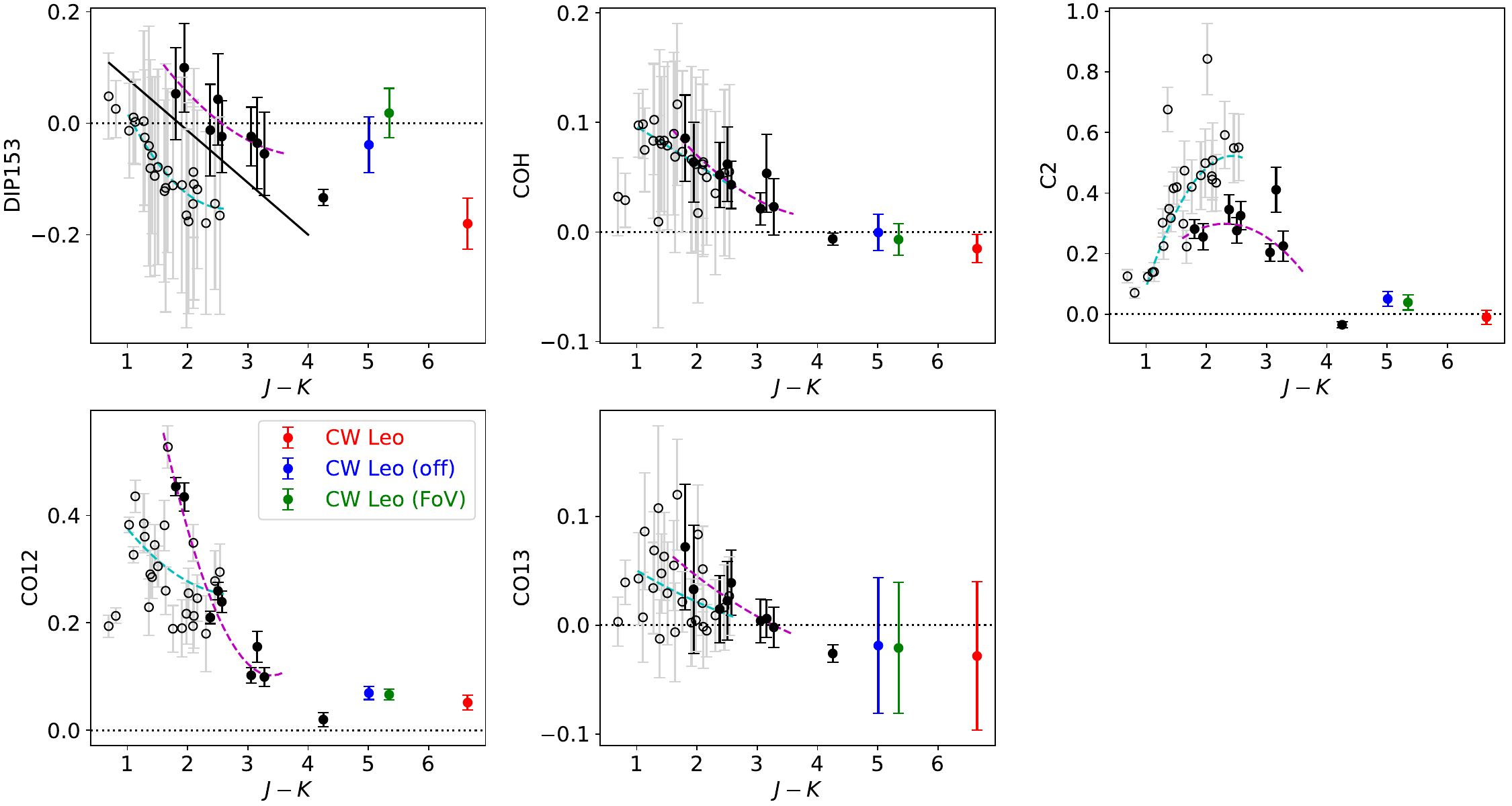}
  \caption{\label{fig:gonn}
    Reproduction of the spectroscopic indices following the method defined
    in \citet{gon16}. See Figure\,\ref{fig:indx} for details, except for
    vertical bars indicating uncertainties propagated from the fluctuation
    in the relevant pseudocontinuum band pass.
  }
\end{figure*}

Corresponding to those derived in this paper and drawn in Figure\,\ref{fig:indx}, the previous spectroscopic indices in \citet{gon16} are reproduced in Figure\,\ref{fig:gonn}. Unlike the continuum fit method employed in this paper, many of the spectroscopic indices in the latter figure have negative values, which would be mistaken as emission. Nevertheless, the two methods display a similarity in the overall trends along $(J-K)$ color, traced by the magenta and cyan dashed lines in the figures showing the fitting results of the template stars: the decreasing trends of DIP153, COH, CO12, and CO13 indices and the maximization of the C2 index near $(J-K)=2.5$. The spectroscopic indices of CW Leo are likely above the extrapolation of the fits for template carbon stars, although they are less outstanding in Figure\,\ref{fig:gonn}.

\begin{figure*} 
  \plotone{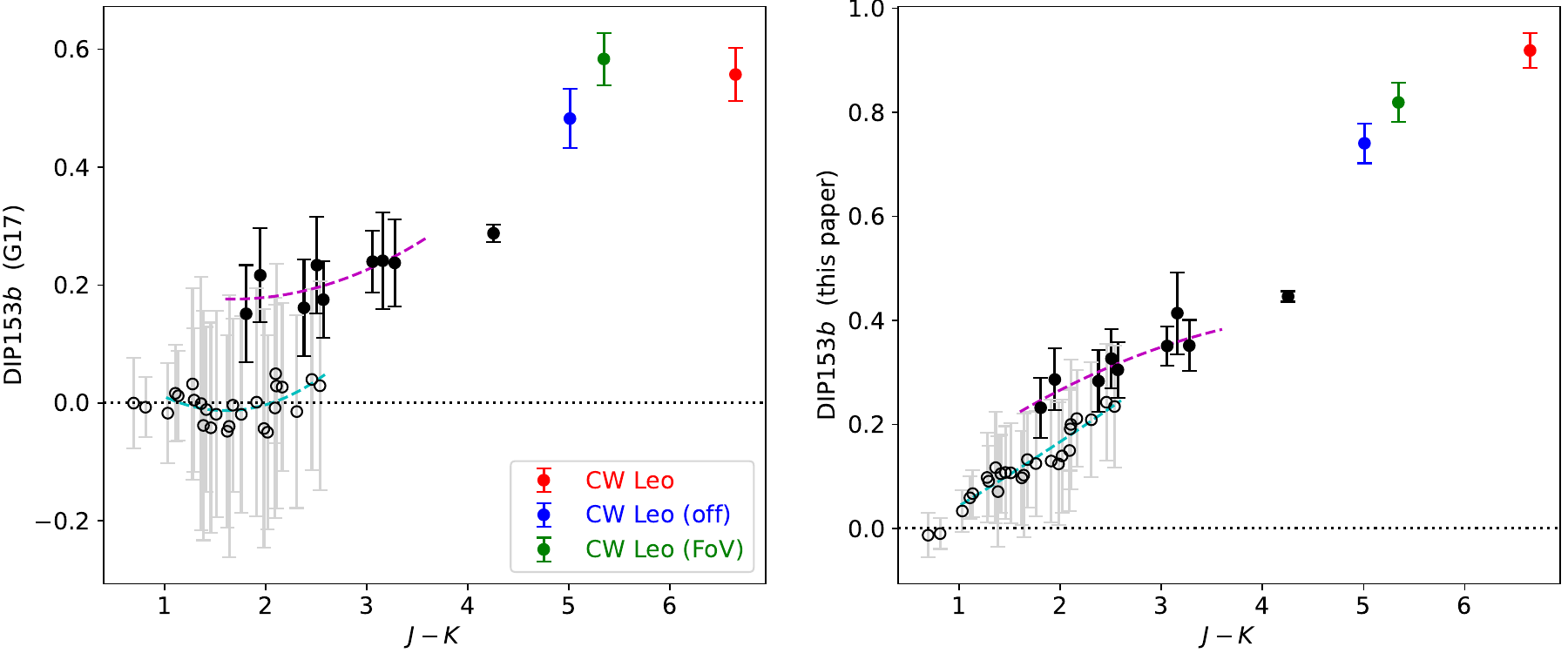}
  \caption{\label{fig:153b}
    A new index DIP153$b$ defined by \citet[G17]{gon17} through an empirical
    fitting for the DIP153 index used in \citet{gon16}, locating open circles
    in the left panel near the zero level. For the right panel, the same
    formula is applied to the DIP153 index used in this paper based on the
    continuum fitting over the $H$ band. See Figures\,\ref{fig:indx} and
    \ref{fig:gonn} for details.
  }
\end{figure*}

\citet{gon16} found that there is a definite separation between two groups of template carbon stars in the graph for the DIP153 index as a function of $(J-K)$ color (see the first panel of Figure\,\ref{fig:gonn}). The stars marked with filled and open circles, respectively, were labeled as stars {\it with} and {\it without} the 1.53\,\micron\ feature, although the numerical values for the index of the 1.53\,\micron-{\it feature-less} stars vary across the color index.
\citet{gon17} introduced a new spectroscopic index ${\rm DIP153}b={\rm DIP153}-0.132\,(1.06-(J-K))$, assigning a value of nearly 0 for the 1.53\,\micron-{\it feature-less} stars in their samples (see the left panel of Figure\,\ref{fig:153b}). This new index for the 1.53\,\micron-{\it present} stars tends to be near a constant value of $\sim0.2$ with a weakly increasing tendency along $(J-K)$ color in the left panel of Figure\,\ref{fig:153b}.
With the definition of continuum used in this paper, the new DIP153$b$ index is shown in the right panel of Figure\,\ref{fig:153b}, exhibiting increasing trends for both categories of stars marked with open and filled circles. These tendencies seem to be valid, despite the large fluctuation in the spectra of the template 1.53\,\micron-{\it feature-less} stars.

In any case, the spectroscopic index for the 1.53\,\micron\ absorption feature of CW Leo is outstanding relative to all template carbon stars.

\end{document}